\documentclass[12pt]{article}
\usepackage{cite,graphicx,amsmath,amssymb}
\usepackage{graphicx}
\usepackage{multicol}
\usepackage{xcolor}
\usepackage{slashed}
\usepackage[normalem]{ulem}

\newcommand{\be}{\begin{equation}}
\newcommand{\ee}{\end{equation}}
\newcommand{\bea}{\begin{eqnarray}}
\newcommand{\eea}{\end{eqnarray}}
\begin{document}

\thispagestyle{empty}
\begin{flushright}
CCTP-2019-2, ITCP-IPP 2019/2, MAD-TH-19-02
\end{flushright}

\vfil
\vspace*{-.2cm}
\noindent

\vspace*{1.2cm}

\begin{center}

{\Large\bf Understanding KKLT \\[.3cm] from a 10d perspective}
\\[2cm]

{\large Yuta Hamada$^{1}$, Arthur Hebecker$^{2}$, Gary Shiu$^{3}$ and Pablo Soler$^{2}$}

\vfil

{\it
${}^{1}$ {Crete Center for Theoretical Physics, Institute for Theoretical and Computational Physics, Department of Physics, University of Crete, P.O. Box 2208, 71003 Heraklion, Greece}\\
 ${}^{2}$ Institute for Theoretical Physics, University of Heidelberg, 
Philosophenweg 19,\\ D-69120 Heidelberg, Germany\\
${}^{3}$ Department of Physics, University of Wisconsin, Madison, WI 53706, USA
}

\vspace{.3cm}
\large{04 February 2019}
\\[1.6cm]

{\bf Abstract}
\end{center}
Some of the most well-celebrated constructions of metastable de Sitter vacua from string theory, such as the KKLT proposal, involve the interplay of gaugino condensation on a D7-brane stack and an uplift by a positive tension object. These  constructions have recently been challenged using arguments that rely on the trace-reversed and integrated 10d Einstein equation. We give a critical assessment of such concerns. We first relate an integrated 10d Einstein equation to the extremization condition for a 10d-derived 4d effective potential. Then we argue how to obtain the latter from a 10d action which incorporates gaugino condensation in a (recently proposed) manifestly finite, perfect-square form. This effective potential is consistent with 4d supergravity and does not present obstacles for an uplifted minimum. Moreover, within standard approximations, we understand the uplift explicitly in one of the popular versions of the integrated 10d equation. Our conclusion is that  
de Sitter constructions of the KKLT type cannot be dismissed simply based on the integrated 10d equations considered so far.

\newpage
\section{Introduction}\label{Sec:introduction}

To establish convincingly that string theory may provide metastable de Sitter vacua remains an important challenge. While seminal ideas to construct such vacua have been proposed (the leading candidates being \cite{Kachru:2003aw, Balasubramanian:2005zx}), attempts for explicit constructions vary in their levels of details. It is fair to say that a fully explicit de Sitter vacuum in string theory remains much to be desired. Even though the ingredients used seem reasonable, the devil may be in the details of implementing them in concrete string compactifications.
With recent Swampland conjectures questioning the existence of controlled de Sitter vacua on general grounds~\cite{Obied:2018sgi,Ooguri:2018wrx}\footnote{See also~\cite{Garg:2018reu} for related conjectures.}, this task becomes even more pressing. 

The goal of this paper is simple. Using our recent proposal for a perfect-square structure of the 10d description of gaugino condensation~\cite{Hamada:2018qef} (see also \cite{Kallosh:2019oxv}), we give a critical assessment of the concerns about the KKLT construction raised in \cite{Moritz:2017xto} and discussed further in~\cite{Gautason:2018gln}. (For more on the debate about de Sitter vacua in string theory, see e.g.~\cite{Bena:2009xk,McOrist:2012yc,Dasgupta:2014pma,Bena:2014jaa,Quigley:2015jia,Cohen-Maldonado:2015ssa,Junghans:2016abx,Moritz:2017xto,Sethi:2017phn,Danielsson:2018ztv,Moritz:2018sui,Cicoli:2018kdo,Kachru:2018aqn,Kallosh:2018nrk,Akrami:2018ylq,Moritz:2018ani,Bena:2018fqc,Kallosh:2018psh,Gautason:2018gln,Armas:2018rsy,Hebecker:2018vxz,Heckman:2018mxl,Junghans:2018gdb,Kallosh:2019axr,Heckman:2019dsj}.) Concretely, our approach involves the following conceptually simple steps:

First, we note that one of the main pillars of the criticism raised in~\cite{Moritz:2017xto} is an analysis (in the spirit of the Maldacena-Nunez no-go theorem \cite{Maldacena:2000mw}) of the integrated trace-reversed 10d Einstein equation. We begin by arguing that (a generic version of) the trace-reversed integrated Einstein equation is equivalent to the condition that an appropriately defined 4d effective potential is extremized. The 4d curvature is then determined, in a very conventional way, by the value of this 4d potential at the extremum. This is indeed not surprising since, by tracing the 10d Einstein equations and integrating them over the compact space, one basically limits the attention to the extremization in two variables: the 4d and the 6d scale factors. The corresponding equations of motion are just the Einstein equation for the 4d curvature scalar and the extremality of the compactification with respect to the total volume~\cite{Giddings:2005ff}. 

Thus, we are left with the task of extremizing the 4d effective potential. To keep the spirit of doing everything explicitly from an ultraviolet completion point of view, we would not take the 4d supergravity formula for this potential in \cite{Kachru:2003aw} for granted. Instead, we obtain this potential (the `10d-derived 4d effective potential') by integrating the 10d action including the gaugino condensate~\cite{Derendinger:1985kk,Dine:1985rz}, its coupling to the 3-form flux \cite{Camara:2004jj,Baumann:2006th,Koerber:2007xk,Baumann:2009qx,Baumann:2010sx,Dymarsky:2010mf} and the uplifting positive-tension brane over the compact space. We do so using the results of GKP~\cite{Giddings:2001yu}, including the warping effects required for the leading-order no-scale structure, and the perfect-square-type gaugino condensate term of~\cite{Hamada:2018qef} (motivated by related structures in M-theory~\cite{Horava:1995qa,Horava:1996ma,Horava:1996vs,Nilles:1997cm,Mirabelli:1997aj}). The latter is particularly crucial as it is free of the UV regularization dependence that plagued the analysis of~\cite{Moritz:2017xto, Gautason:2018gln}. This perfect-square structure is also consistent with known 4d and 10d SUSY constraints, especially the quartic gaugino term (see~\cite{Kallosh:2019oxv} for a discussion and more references). Through these procedures, we find a potential that agrees with the 4d SUGRA expectations. 

We go on to use a slightly different form of the integrated trace-reversed 10d Einstein equations (the one arising after implementation of the 5-form Bianchi identity)~\cite{Giddings:2001yu,deAlwis:2003sn,Moritz:2017xto, Gautason:2018gln} to check for the uplift explicitly. Within standard approximations and using our perfect-square form of the gaugino-condensate term, we find agreement with the previously obtained positive potential. Thus, the variants of the integrated trace-reversed 10d Einstein equations considered so far (equivalently, the `10d derived 4d effective potential') pose no obstructions to a de Sitter uplift.

Of course, this type of analysis does not replace a full 10d solution. In fact, it is not obvious with which degree of explicitness the latter can be obtained (although an existence proof may be a realistic goal for the future)\footnote{To draw on the analogy made in \cite{Polchinski:2015bea}, while determining the precise energy spectrum of an atom would require the full machinery of quantum mechanics, one may as a first step try to argue for the existence of bound states.}. However, what we have achieved independently of such a full solution is the clear demonstration that one cannot dismiss KKLT on the basis of a simplified argument using just the known forms of integrated trace-reversed equations: The latter are consistent with the 4d effective approach (as defined in the text) and this, in turn, is consistent with known 4d SUGRA results. Thus, in absence of a fully explicit 10d solution, KKLT appears to survive the level of scrutiny that can so far be implemented.

\section{The 10d and the 4d approaches to 4d curvature.}\label{trrg}

\subsection{General setting}

Two approaches can be taken to obtain the value of the on-shell 4d curvature ${\cal R}_4$ of string compactifications. From a 4d effective point of view, as is standard, one finds the minimum $V_0$ of the potential for the moduli fields, and plugs its value into the 4d trace-reversed Einstein equations, from which one immediately reads ${\cal R}_4=4V_0$. Alternatively, one can use the 10d trace-reversed Einstein equations  directly in the higher-dimensional theory and take the trace over its non-compact components to obtain ${\cal R}_4$~\cite{Maldacena:2000mw,deAlwis:2003sn}. Of course, both approaches should lead to the same result~\cite{Giddings:2005ff,Danielsson:2009ff,Danielsson:2011au}.

It will be useful for the rest of our paper to explicitly derive this equivalence in a generic, warped setting (see also the more detailed treatment of~\cite{Giddings:2005ff}). We start with the $(4+n)$-dimensional action
\be\label{action}
S=\int d^4x \, d^ny\,\sqrt{- G}\left(\frac{1}{2}{\cal R}_{G}+{\cal L}\right)\,.
\ee
The metric $G$ takes the warped form
\be
ds^2=\Omega(y)^2\,\eta_{\mu\nu}\,dx^\mu dx^\nu+g_{mn}dy^m dy^n\,,
\label{wme}
\ee
where $\eta_{\mu\nu}$ is a maximally symmetric 4d metric independent of $y$, and $g_{mn}$ is a generic $n$-dimensional metric. It is easy to separate from the total warped curvature ${\cal R}_{G}$ the unwarped 4d one ${\cal R}_\eta$. In fact, one can see that~\cite{Maldacena:2000mw}:
\be
{\cal R}_{G}=\Omega^{-2} {\cal R}_\eta+ {\cal R}_G^{(n)}+\Omega^{-4}\times\text{(total-derivative)}\label{reta}
\ee
where ${\cal R}_G^{(n)}=G^{mn}({\cal R}_G)_{mn}$. The last term clearly does not contribute to the action. Hence, (\ref{action}) becomes
\be
S=\int d^4x\,d^ny\,\sqrt{-\eta}\sqrt{g}\left[  \Omega^{2}\frac{1}{2}{\cal R}_\eta+ \Omega^4\left(\frac{1}{2}{\cal R}_G^{(n)}+{\cal L}\right)\right]\,.
\ee
One can then integrate over the inner space to obtain the 4d action
\be\label{4daction}
S=\int d^4 x\sqrt{-\eta}\left[\frac{{\cal V}_n}{2} {\cal R}_\eta +\int d^n y \,\sqrt{g}\, \Omega^{4} \left(\frac{1}{2}{\cal R}_G^{(n)}+{\cal L}\right)\right]\,,
\ee
where we have defined the `inner volume'
\be\label{eq:volumedefinition}
{\cal V}_n=\int d^ny\,\sqrt{g}\,\Omega^{2}\,.
\ee

\subsection{The 4d equations}

In a homogeneous background, the 4d Einstein equations derived from~\eqref{4daction} lead to
\be\label{4dEE}
{\cal V}_n{\cal R}_\eta=-4\int d^n y \,\sqrt{g}\, \Omega^{4} \left( \frac{1}{2}{\cal R}_G^{(n)}+{\cal L}\right)\,.
\ee
The second 4d equation that we need comes from the extremization of the action with respect to scalings of the inner metric $g_{mn}$. If we denote the corresponding length-scale modulus by $R$, the requirement that the 4d action is extremized in $R$ then reads\footnote{
Notice 
that $R$ does not have to coincide with the flat direction, usually denoted by $T$, in GKP-type compactifications. While $R$ corresponds to scalings of $g_{mn}$ only, $T$ involves simultaneous rescalings of the warp factor $\Omega$. Of course, both the equations of motion for  $T$ and for $R$ should be satisfied on-shell. In other words, strictly speaking $V_4=V_4(R,T)$ and is stationary in both variables in case of a solution. The difference between $R$ and $T$ is significant only when contributions are localized to strongly warped regions where $\Omega \ll 1$, in particular for $\overline{D3}$ branes in the KKLT scenario.}
\be
\frac{\partial V_4(R)}{\partial R}=0 \qquad \Longleftrightarrow\qquad \int d^ny\,\sqrt{g}\,g^{mn}\frac{\delta S}{\delta g^{mn}}=0\,.
\ee 
Here $V_4$ is the 10d-derived 4d effective potential in Einstein-Frame, which is explicitly given by the right hand side of (\ref{4dEE}), after division by $4{\cal V}^2_n$.

The extremization condition in $R$ or $g_{mn}$ can be written as
\be\label{4deom}
\int d^ny\,\sqrt{g}\,\Omega^{4}\left[\frac{n+2}{2}\,\frac{1}{2}{\cal R}_G^{(n)}+ \frac{n}{2}{\cal L}+g^{mn}\frac{\delta {\cal L}}{\delta g^{mn}}\right]=0\,.
\ee
We want to show that eq.~\eqref{4dEE}, evaluated at a solution of~\eqref{4deom}, results in a curvature that can also be obtained from the (traced) $(4+n)$-dimensional trace-reversed Einstein equations. For that, we will find it useful to combine~\eqref{4dEE} and~\eqref{4deom} in an expression where ${\cal R}_G^{(n)}$  does not appear:
\be\label{eq:4dresult}
{\cal V}_n{\cal R}_\eta=\frac{8}{n+2}\int d^n y \,\sqrt{g}\, \Omega^{4} \left( -{\cal L}+g^{mn}\frac{\delta {\cal L}}{\delta g^{mn}}\right)\,.
\ee
Let us stress that this last expression holds only on-shell, i.e. when evaluated at a solution of~\eqref{4deom}.

\subsection{10d equation}\label{sec:10dgeneral}
Instead of dimensionally reducing to 4d, one can study the curvature using directly the higher-dimensional Einstein equations. Their trace-reversed version in $(4+n)$-dimensions reads
\be
\left({\cal R}_{G}\right)_{MN}=T_{MN}-\frac{1}{n+2}\,T\,{G}_{MN}\,.
\ee
Tracing this over the non-compact components one finds
\be\label{10dEE}
({\cal R}_G)_4\equiv G^{\mu\nu}\left({\cal R}_{G}\right)_{\mu\nu}=\frac{1}{n+2}\left[(n-2)T_4-4T_n\right]\,,
\ee
where
\begin{eqnarray}
T_4&=&G^{\mu\nu}T_{\mu\nu}=4\,{\cal L}-2G^{\mu\nu}\frac{\delta {\cal L}}{\delta G^{\mu\nu}}=4\,{\cal L}\nonumber\\
T_n&=&G^{mn}T_{mn}=n\,{\cal L}-2G^{mn}\frac{\delta {\cal L}}{\delta G^{mn}}\,.
\end{eqnarray}
In the first line we used the fact that in a homogeneous background, the Lagrangian does not have a non-trivial tensor structure in the non-compact direction. 
The definition of the energy momentum tensor is
\be
T_{\mu\nu}\equiv-{2\over\sqrt{-G}}  {\delta S\over \delta G^{\mu\nu}},
\quad
T_{mn}\equiv-{2\over\sqrt{-G}}  {\delta S\over \delta G^{mn}}.
\ee

The combination appearing on the right hand side of (\ref{10dEE}) is
\be
(n-2)T_4-4T_n=8\left(-{\cal L}+G^{mn}\frac{\delta {\cal L}}{\delta G^{mn}}\right)\,.
\ee
As before, the curvature term on the left hand side of (\ref{10dEE}) can be expressed in terms of~${\cal R}_\eta$:
\be\label{eq:totalderivative}
({\cal R}_G)_4=\Omega^{-2} {\cal R}_\eta +\Omega^{-4}\times\text{(total-derivative)}\,.
\ee
Since the last term vanishes upon integration over the internal space, we may rewrite~\eqref{10dEE} in the form
\be\label{10d}
\int d^ny\,\sqrt{g}\,\Omega^{2}\,{\cal R}_\eta=\frac{8}{n+2}\int d^ny\,\sqrt{g}\,\Omega^{4}\left(-{\cal L}+g^{mn}\frac{\delta {\cal L}}{\delta g^{mn}}\right)\,.
\ee
But this is precisely the same as~\eqref{eq:4dresult}, which we obtained previously by combining the 4d Einstein equations and the equation of motion for the volume modulus. This confirms~\cite{Giddings:2005ff,Danielsson:2009ff,Danielsson:2011au} that the 10d and 4d approaches are equivalent in determining the on-shell 4d curvature ${\cal R}_\eta$.\footnote{
This 
equivalence requires that, when varying the action, the dependence of every term in ${\cal L}$ on the compactification volume is properly taken into account. In particular, this applies to non-perturbative terms $\sim e^{-T}$ depending on the Kahler modulus $T$. Such a non-local dependence of 10d lagrangrian terms on the compact volume may look awkward, but it is crucial in pursuing a 10d approach.
}

The 10d approach has often been used to provide no-go theorems for the construction of de Sitter vacua in string theory~\cite{Maldacena:2000mw}. A popular way of writing~\eqref{10d} in this approach is roughly as
\be
{\cal V}\,{\cal R}_\eta=\int d^6y\sqrt{g}\,\left(-2\Omega^4\,\Delta\right)\qquad\mbox{with}
\qquad\Delta=\frac{1}{4}(T_m^m-T_\mu^\mu)\,.\label{vr1}
\ee
Here we have set $n=6$, as appropriate for the Calabi-Yau case, but maintained the convention $M_{P,\,10}=1$ for simplicity. As before, the internal volume ${\cal V}$ is defined with a factor $\Omega^2$ under the integral~\eqref{eq:volumedefinition}, but we keep the generic notation ${\cal V}$. Indeed, volume integrals are always dominated by the unwarped region ($\Omega\simeq 1$) and for our leading order analysis it will never be of importance with which factor of $\Omega$ a volume integral is calculated in any particular equation.

A slight variation of this relation (representing conceptually still the same physics) can be derived in the context of type IIB compactifications. It has in particular been used in~\cite{Moritz:2017xto,Gautason:2018gln} to argue against uplifting to de Sitter:
\be
{\cal V}\,{\cal R}_\eta=\int d^6y \sqrt{g}
\left( - a\,|\partial \Phi^-|^2-b\,\Omega^8\,\frac{|G_3^-|^2}{\mbox{Im}(\tau)}-2\Omega^8 \Delta^{other}\right)\,.\label{vr2}
\ee
Here $G_3^-$ is the imaginary anti-selfdual part of the complexified 3-form flux, $\Phi^-=\Omega^4-\alpha$ is a particular combination of the warp factor and the 4-form potential ($\tilde{F}_5=\mbox{$(1+*)$}\,d\alpha\wedge \,d\,V\!ol_4$), and $a,b$ are numerical coefficients. Furthermore, $\Delta^{other}$ is built, in analogy to $\Delta$ of (\ref{vr1}), from the energy momentum tensor of other contributions. Those are in particular terms involving the gaugino-condensate and the $\overline{D3}$-branes required for the uplift.

Now, the arguments of~\cite{Moritz:2017xto,Gautason:2018gln} go roughly as follows: The first two terms on the right hand side of (\ref{vr2}) are mainfestly non-positive and, in fact, exactly zero before gaugino condensation. Hence, in the case of an AdS vacuum based on a gaugino condensate, the last term is expected to be negative as it represents the main new effect (more in Section \ref{Sec:cancellation}). Adding an $\overline{D3}$ brane as an uplift, one naively does not expect to achieve de Sitter since (as one easily checks) an anti-brane with its localised positive energy density contributes positively to $\Delta^{other}$. The potentially negative change of $\Delta^{other}$ from the backteaction of the 
$\overline{D3}$ on the gaugino condensate was argued to be not strong enough.

We will argue the opposite using both forms of the integrated 10d Einstein equation. Indeed, since we showed that (\ref{vr1})
 is equivalent to the 4d potential-energy-based analysis, this implies that if we can rederive the KKLT energetics from 10d, then the change of $\Delta$ induced by the gaugino condensate (backreacting to the $\overline{D3}$ brane) must be strong enough to provide de Sitter. In addition, we will use (\ref{vr2}) to calculate the 4d curvature explicitly from $\Delta$. We will also explain in some detail the differences between the two forms of the integrated 10d Einstein equation. While (\ref{vr1}) is more suitable for showing the equivalence of the 10d and 4d approaches, we will discuss why (\ref{vr2}) is more easily usable for an explicit consistency check in the strongly warped case.

Following the outline given above, in the next section we will derive (without using 4d supergravity) the 4d effective theory of the KKLT construction starting from a 10d theory, in particular making use of the quartic gaugino action previously proposed by us.

\section{Deriving the 4d effective potential from 10d}\label{Sec:4d_potential}

\subsection{Including the gaugino terms in the 10d action}

Following~\cite{Giddings:2001yu}, our starting point is the type IIB action in the 10d Einstein frame:
\begin{align}\label{eq:IIB_action}
S=
{1\over2\kappa_{10}^2}
	\bigg[
	&\int \left({\cal R}_{10}*1-{ig_s\over4} C_4\wedge G_{3}\wedge \overline{G}_{3}\right)
	-{g_s^2\over2}\left|d\tau\right|^2-{1\over4}\left|F_5\right|^2	
	&-{g_s\over2}\left|G_{3}\right|^2
	\bigg]
+S_{\rm loc}\,.
\end{align}
Here $S_{\rm loc}$ is the action of localized objects (D-branes and O-planes). Fermions will mostly be irrelevant and have been suppressed. However, crucially, the D7-brane gauginos will play a central role and have to be included. We write their action in shorthand, focussing on the interaction between gauginos and 3-form flux:
\be
S_{\lambda\lambda}=c\sqrt{g_s}\int_XG_3\wedge \Omega_3\,\overline{\lambda}\overline{\lambda}\,\delta_{D7}\,+\,\mbox{c.c.}\,+\,\,\mbox{kinetic and $\lambda^4$ terms}\,.
\ee

Now, the key proposal of~\cite{Hamada:2018qef} is that the $|G_3|^3$ term, the $G_3\,\lambda\lambda$ term and the gaugino quartic term combine in total square form, as expected from the heterotic case. Concretely, the proposal reads
\begin{align}\label{2bw}
S=
{1\over2\kappa_{10}^2}
	\bigg[
	&\int \left({\cal R}_{10}*1-{ig_s\over4} C_4\wedge G_{3}\wedge \overline{G}_{3}\right)
	-{g_s^2\over2}\left|d\tau\right|^2-{1\over4}\left|F_5\right|^2	
	\nonumber\\
	&-{g_s\over2}\left|G_{3}-P\left(
	c\,{\lambda\lambda\over\sqrt{g_s}}
	\,\overline{\Omega}_3\,\delta_{D7}\right)\right|^2
	\bigg]
+S_{\rm loc}+\cdots,
\end{align}
where we have absorbed the gaugino kinetic term in $S_{\rm loc}$ for ease of notation. Here $P$ is a projection operator on closed forms. Making use of the uniqueness of the Hodge decomposition into harmonic, exact and coexact piece, it removes the coexact contribution (the harmonic piece, which is also closed, being preserved as well). Explicitly, following the torus example discussed in detail in~\cite{Hamada:2018qef}, one has
\be
P(\overline{\Omega}_3\,\delta_{D7})=\frac{\overline{\Omega}_3}{A}+\mbox{exact}\,.\label{pdef}
\ee
In other words, the harmonic part is proportional to $\overline{\Omega}_3$ (see Appendix) and suppressed by the brane-transverse volume $A \sim {\cal V}/{\cal V}_{D7}$. The latter is required by the corresponding scaling of the $\delta$-function $\delta_{D7}$. 

Furthermore, we allow for some 3-form flux to be present in the background. In other words, we let
\be
G_3=G_3^{(0)}+\delta G_3\,,\label{g3def}
\ee
where $G_3^{(0)}$ is the standard combination of integer, harmonic $F_3$ and $H_3$ forms representing the flux and $\delta G_3$ is trivial in cohomology. Restricting ourselves to gaugino zero modes, such that $\lambda\lambda$ is a constant prefactor, and using (\ref{pdef}) and (\ref{g3def}) in (\ref{2bw}), one obtains
\begin{align}
S={1\over2\kappa_{10}^2}
	\bigg[
	&\int \left({\cal R}_{10}*1-{ig_s\over4} C_4\wedge G_{3}\wedge \overline{G}_{3}\right)
	-{g_s^2\over2}\left|d\tau\right|^2-{1\over4}\left|F_5\right|^2	
	\nonumber\\
	&-{g_s\over2}\left|G_{3}^{(0)}-c{\lambda\lambda\over\sqrt{g_s}A}\overline{\Omega}_3\right|^2
	\bigg]
+S_{\rm loc}\,.\label{acts}
\end{align}
Here $\delta G_3$ has adjusted to cancel the exact piece of (\ref{pdef}), such that only the harmonic flux part $G_3^{(0)}$ and the gaugino-induced contribution $\sim \overline{\Omega}_3$ are left inside the total square. The field strength $G_3$ in the first line should be understood as the on-shell value of \eqref{g3def}. We do not write it explicitly in order to avoid the complicated expression. 

Next, we appeal to GKP~\cite{Giddings:2001yu} to claim that, if the gaugino condensate vanishes ($\langle \lambda\lambda\rangle = 0$), 
the action of (\ref{acts}) leads to an identically vanishing 4d effective potential. Macroscopically, this is simply a result of the no-scale structure of the Kahler moduli Kahler potential. As a consequence, if one allows for a non-zero gaugino condensate, then the 4d effective potential will follow from (\ref{acts}), restricted to the terms which involve $\lambda\lambda$:
\be
S_{\lambda\lambda}={1\over2\kappa_{10}^2}\int\left[
\frac{g_s}{2}G_3^{(0)}\wedge\overline{\left(c{\lambda\lambda\over\sqrt{g_s}A}\overline{\Omega}_3\right)}+\mbox{c.c}-\left|c{\lambda\lambda\over\sqrt{g_s}A}\overline{\Omega}_3\right|^2\right]\,.
\label{gte}
\ee
Before turning this into a 4d effective potential, we insert a brief subsection justifying the last step.

\subsection{Treatment of $\left|G_3^{(0)}\right|^2$ and gaugino-induced internal curvature}\label{Sec:cancellation}

The $\left|G_3^{(0)}\right|^2$ term should not appear in the 4d effective potential, reflecting the no-scale Kahler potential. The argument is roughly as follows. From the equation of motion/Bianchi identity of 5-form flux, we have
\begin{align}
dF_5 = {i\over2} g_s G_3\wedge\overline{G}_3 + 2\kappa_{10}^2 T_3 \rho_{3}^{\rm loc}.
\end{align}
Integrating over the internal space, we can get 
\begin{align}
g_s \int_6 \left|G_3^{(0)}\right|^2 = -4\kappa_{10}^2 T_3 \left(\int d^6y\,\rho_{3}^{\rm loc} \sqrt{-g}\right).
\end{align}
For a D3, D7 and O3 system, this cancels the tension terms at the leading order of $\alpha'$.\footnote{
Jumping 
somewhat ahead, we also note that the inclusion of a $\overline{D3}$-brane enhances this tension term, which results in the prefactor $2$ of the contribution to the potential from $\overline{D3}$-brane~\cite{Kachru:2003aw}.}

This is, however, not the end of the story: The background flux $G_3^{(0)}$ and the D-branes/O-planes induce both warping (i.e.~a non-vanishing internal-space Ricci scalar) as well as a non-trivial $F_5$-form field strength. It is one of the central and widely known results of~\cite{Giddings:2001yu} that all these effects combine in just the right way to allow for a solution of the 10d Einstein equations. In our context this means that the corresponding terms in (\ref{acts}) are locally non-zero  but vanish upon integration over the compact space. We emphasize in particular that non-zero warping is essential to allow for a separation between the positive and negative tension sources in 10d. Thus, when going from (\ref{acts}) to (\ref{gte}), we do not neglect warping but make use of the result of~\cite{Giddings:2001yu} which  shows that its integral effect vanishes.

However, since we are aiming at the value of the 4d effective potential in a minimum, corresponding to a 10d solution, we have to be concerned about additional internal-space curvature effects and the backreaction of other fields induced by a non-zero value of the $\lambda\lambda$ terms in (\ref{acts}). Similar effects can come from the $\overline{D3}$ uplift to be included below. To convince ourselves that they are negligible, we rewrite (\ref{acts}) symbolically as
\be
S=\int_{10}\sqrt{-G}\,\left({\cal R}_\mu{}^\mu+{\cal R}_m{}^m+{\cal L}+\delta{\cal L}\right)\,,\label{sact}
\ee
see also (\ref{reta}). Furthermore, we parameterize the metric as
\be
G=(\eta,\Omega,{\cal V},g)\,,
\ee
where $\eta$ is the maximally symmetric (AdS, dS or Minkowski) 4d metric, $\Omega$ is the warp factor (normalized to unity in the bulk), ${\cal V}$ the compact volume\footnote{
We 
focus on the case of a single Kahler modulus. The generalisation is straightforward.
}
and $g$ the internal metric normalised to unit volume.  We know from GKP that a solution with ${\cal R}_\mu{}^\mu=0$ exists for $\delta{\cal L}=0$. 

Now we turn on a perturbation which we call $\delta {\cal L}$ corresponding to new terms in the lagrangian involving $\lambda\lambda$ or the $\overline{D3}$ brane or both. Crucially, by $\delta {\cal L}$ we mean {\it not} the total change of the lagrangian but only the new, explicitly added terms (after the naive UV divergences introduced by the gaugino condensate are regulated in ${\cal L}$). We assume that a new, modified solution exists. 

Most naively, the new solution will be characterized by a new, in general non-zero 4d curvature ${\cal R}_\eta$ and hence a non-zero integral of ${\cal R}_\mu^\mu$. By the 4d Einstein equations, this perturbation of the old, vanishing ${\cal R}_\mu^\mu$ is linear in $\delta {\cal L}$. Our proposal is to calculate the 4d curvature by using just $\delta {\cal L}$ on the r.h. side of the 4d Einstein equations, i.e. without worrying about any possible effect from a modified ${\cal R}_m^m$ or ${\cal L}$ in (\ref{sact}). To justify this, we will argue that these effects are ${\cal O}(\delta {\cal L}^2)$. 

Any modification of the ${\cal R}_m^m$ and ${\cal L}$ must come from the perturbations $\delta \eta$, $\delta {\cal V}$, $\delta \Omega$ and $\delta g$, each of which is linear in $\delta {\cal L}$. First, $\delta \eta$ is irrelevant since it appears only as a prefactor and the integral over ${\cal R}_m^m+{\cal L}$ was zero originally, as we know from GKP. Second, ${\cal V}$ is a modulus to begin with and we can hence set $\delta {\cal V}=0$ by choosing which solution we consider to be the unperturbed one. Thus, we only have to deal with $\delta \Omega$ and $\delta g$ and their effect on $S$. The two quantities $\Omega$ and $g$ are fixed to start with. In other words, by the stationarity of the action on the GKP solution, $S$ is quadratic in perturbations $\delta \Omega$ and $\delta g$ at $\delta {\cal L}=0$. Turning on a small perturbation $\delta {\cal L}$, they react linearly, $\delta \Omega\sim \delta g \sim \delta{\cal L}$. Hence, in the variation of the action induced by $\delta {\cal L}$, their effect is subleading:
\be
\delta \left[ \int_{10}\sqrt{-G}\,\left({\cal R}_m{}^m+{\cal L}\right)\right]
\sim \alpha\,(\delta\Omega)^2 +\beta\,(\delta g)^2 \,\,\sim\,\, (\delta {\cal L})^2\,.
\ee
This ends our excursion aimed at justifying (\ref{gte}). We emphasize again that the same argument justifies our analogous treatment of the $\overline{D3}$ potential (without backreaction effects) further down.

Two important remarks can be made at this point: First, if the perturbation is not localized in a strongly warped region, then we may assume ${\cal R}_m^m=0$ which is moreover a (Calabi-Yau) solution to the matter-free Einstein equations. It is thus by itself quadratic in $\delta g$ and $\delta \Omega$. The argument above then shows that $\int {\cal L}$ is also by itself quadratic in $\delta g$ and $\delta \Omega$. One may hence calculate the crucial quantity $\Delta$ from $\delta {\cal L}$ alone.

Second, by contrast to the above, if a perturbation $\delta {\cal L}$ is localized in a strongly warped region, then only the combination $\int ({\cal R}_m^m+{\cal L})$ is stationary and hence quadratic in $\delta {\cal L}$. But $\Delta$ is calculated from the lagrangian alone and thus it is not clear that one may disregard backreation when calculating it in a perturbed situation. This is the reason that further down, in Sect.~\ref{Sec:uplift_trace-reversed}, it will be advantageous to use the version (\ref{vr2}) rather than (\ref{vr1}) of the integrated Einstein equations to work out the 4d curvature explicitly. Indeed, in (\ref{vr2}) the backreaction effects are automatically subleading since the first two terms under the integral are quadratic in the fields and zero in the GKP solution.

Finally, one may use a trick to circumvent the problem above and still obtain the 4d curvature from (\ref{vr1}), even in the case of an uplifting $\overline{D3}$ in a warped throat: One may simply view the warped throat as a confining 4d gauge theory localized in an otherwise unwarped Calai-Yau. The warping, including the nontrivial dependence of the warping on the volume modulus, is then encoded in an ad-hoc numerical prefactor and an ad-hoc $T$-dependence of the uplifting energy momentum tensor. It turns out that the operator $T(\partial/\partial T)-1$ generating $\Delta$ (cf.~(\ref{eq:deltaformula})) exactly annihilates the $\overline{D3}$ action which then does not contribute to $\Delta$. This corresponds, in the approach of (\ref{vr1}), to the $\Omega^8$ suppression of strongly warped regions present in (\ref{vr2}). These comment will become more transparent after the discussion of Sect.~\ref{Sec:trace_reversed}.

\subsection{4d scalar potential before the uplifting}\label{Sec:SUSY_AdS}

We now continue by integrating (\ref{gte}) over the compact space, rescaling the 4d metric to go to the 4d Einstein frame, and rescaling the gauginos to comply with standard SUSY normalization. The result does not change qualitatively compared to the toroidal toy model analysed in~\cite{Hamada:2018qef} (cf.~Eq.~(44) therein):
\begin{align}\label{eq:previous_result}
&
{\cal L} \,\,\sim\,\, {\cal R}_4 \,- T\,\text{tr}\left(F_{\mu\nu}F^{\mu\nu}-i\lambda\slashed{D}\bar{\lambda}+\text{c.c.}\right)-{T^2}\,|\lambda\lambda|^2+\left[\frac{\sqrt{g_s}}{A}\,{\bar\lambda\bar\lambda}\,W_0+\text{c.c.}\right]\,.
\end{align}
Here $W_0\equiv\left(G_3^{(0)},\overline{\Omega}_3\right)$ and we reinstated the gravity, gauge and gaugino kinetic terms, which were not part of (\ref{gte}). We have also rescaled $\Omega_3$ to be independent of the volume, such that $W_0$ is only a function of the (by now stabilized) complex structure moduli. Moreover, we are focussing on models with a single Kahler modulus, such that $T\sim {\cal V}_{D7} \sim {\cal V}^{2/3}$ and $A \sim {\cal V}/{\cal V}_{D7}\sim \sqrt{T}$. All ${\cal O}(1)$ factors have been suppressed.

Crucially, we are implicitly integrating out the axion that comes with $T$ using the same logic as~\cite{Kachru:2003aw}. To be precise, the complex superfield $\tau$ is $\tau=T+ic$ with $c$ a $C_4$ axion. This axion appears in an $\exp(iac)$ prefactor with the last term of (\ref{eq:previous_result}) and it is stabilised in a way that minimises the corresponding contribution to the potential. The net effect of this is equivalent to choosing the VEVs of $\lambda\lambda$ and $W_0$ to be real and positive and ignoring the axion. This is what we do from now on.

We now explicitly assume that the gaugino bilinear develops a VEV and write down the corresponding potential:
\begin{align}
V\,\,\sim\,\,
T^2|\langle \lambda\lambda \rangle|^2 
-\left[\langle{\bar\lambda\bar\lambda}\rangle\sqrt{\frac{g_s}{T}}W_0+\text{c.c.}\right].
\end{align}
The vacuum expectation value is~\cite{Kaplunovsky:1993rd,Kaplunovsky:1994fg}
\be\label{eq:gaugino_condensation}
\langle \lambda\lambda \rangle
\sim {\sqrt{g_s}\over T^{3/2}}\,e^{-a T},
\ee
with $a=2\pi/N$ for an $SU(N)$ gauge group. The non-exponential prefactor (which is not essential for the following analysis) is affected by the super-Weyl anomaly and the precise definition of the UV gauge coupling. The result is known to be consistent with supergravity formulae for the action, where the gaugino condensate is traded for a non-zero holomorphic $W\sim \exp(-a\tau)$. With this, one finds 
\be\label{eq:potential_AdS}
V\,\,\sim\,\, g_s\left( {1\over T} e^{-2aT} - {2\over T^2} W_0 e^{-aT} \right)\,,
\ee
in structural agreement with the KKLT superpotential $W=W_0+B e^{-a\tau}$ with $B={\cal O}(1)$ and the Kahler potential $K=-3\ln (\tau+\overline{\tau})$ at large volume. As is well known, this leads to a supersymmetric AdS minimum determined by the extremisation condition of \eqref{eq:potential_AdS}:
\be\label{eq:TAdS}
-T\,e^{-2aT}+W_0\,e^{-aT}=0\,,
\ee
where we need $W_0\ll 1$ to guarantee a controlled solution with $T\gg 1$.

\subsection{Adding $\overline{D3}$-brane}

The tension $T_3$ of an $\overline{D3}$-brane contributes to the 10d action as
\begin{align}
S_{\overline{D3}}=-\int d^4x\,d^6y\,\sqrt{-g} \,T_3 \delta_{\overline{D3}}
=-\int d^4x\,\sqrt{-g_4} \,T_3\,.
\end{align}
Placing this anti-brane in a strongly warped region, the contribution to the potential is suppressed by the fourth power of the warp factor $\Omega\ll 1$, cf.~(\ref{wme}). Moreover, in the strongly warped regime the warping gets diluted with growing volume: $\Omega\sim T^{1/4}\exp(-2\pi K/3 g_s M)$ where $K$ and $M$ are the flux numbers~\cite{Kachru:2003sx}. This leads to the Einstein-frame effective potential
\be\label{eq:antibraneaction}
V_{\overline{D3}}\,\sim\, 2 {\Omega^4 \over T^3} T_3 \,\sim\, 
2 {e^{-8\pi K/3g_s M}\over T^2} T_3
={2\mu_3\over T^2}\,.
\ee
Here $\mu_3$ is the warping-suppressed ($T$-independent) $\overline{D3}$ tension and the factor $1/T^3$ arose from the Weyl rescaling the 4d metric. We also included a factor of two arising from the interplay with the flux background.\footnote{
Recall 
that, given a fixed $D3$ tadpole contribution, we would have to transform some of the flux into a $D3$ brane and let it annihilate with the present $\overline{D3}$ to return to a SUSY Minkowski vacuum. Hence the energetic effect of the anti-brane is twice that of its tension.
}

The full potential now reads 
\be\label{eq:potential_dS}
V\sim 
g_s\left( {1\over T} e^{-2aT} - {2\over T^2} W_0 e^{-aT} \right)+{2\mu_3\over T^2}\,,
\ee
with its minimum determined by the solution to
\be\label{eq:TdS}
-g_s aT\, e^{-2aT}+g_s a W_0\, e^{-aT}-{2\mu_3\over T}=0\,.
\ee
It is well-known and easily seen that, due to the steepness of the exponential functions responsible for the original AdS minimum, the last terms can act as an `uplift' for suitably small $\mu_3$. The value of $T$ is now slightly larger than in the SUSY AdS minimum. In Sect.~\ref{Sec:uplift_trace-reversed}, we will see explicitly that this is consistent with the trace-reversed Einstein equation. This is of course already clear if one relies on the general analysis of Sect.~\ref{trrg}.

\section{Trace-reversed Einstein equation}\label{Sec:trace_reversed}

In Sect.~\ref{trrg} we showed that computing the 4d scalar curvature ${\cal R}_\eta$ from the 10d Einstein equations is equivalent to its computation directly from the 4d effective theory. We then obtained in Sect.~\ref{Sec:4d_potential} the 4d potential of KKLT starting from the action of type IIB string theory supplemented by the four-fermion localized gaugino interactions proposed in~\cite{Hamada:2018qef}. In combination, these results should by themselves help dispel some of the doubts that have been recently raised about the validity of the KKLT scenario from a 10d treatment.\footnote{Of course, we are not establishing 
in full
 the validity of KKLT.
This remains an open and interesting issue. We are only addressing here a particular set of concerns.}

In the following, we supplement this discussion by computing ${\cal R}_\eta$ in the KKLT scenario directly from the 10d Einstein equations. Although this is somewhat redundant, it may help clarify the origin and resolution of some of the mentioned criticisms.

\subsection{Derivation of the on-shell potential from the trace-reversed Einstein equation}

The 10d analysis of Sect.~\ref{sec:10dgeneral} showed that the on-shell 4d curvature ${\cal R}_\eta$ follows from the 10d Einstein equations as
\be\label{eq:curv10d}
{\cal V}_6\,{\cal R}_\eta=\int d^ny\,\sqrt{g}\,\Omega^{4}\left(-{\cal L}+g^{mn}\frac{\delta {\cal L}}{\delta g^{mn}}\right)=-2\int d^6y\,\sqrt{g}\,\Omega^{4}\Delta
\ee
where we have introduced the common notation
\be\label{eq:Delta}
\Delta =\frac{1}{4}\left(T^m_m-T^\mu_\mu\right)= \frac{1}{2}\left({\cal L}-g^{mn}\frac{\delta {\cal L}}{\delta g^{mn}}\right)\,.
\ee

Our goal is to display how different terms in the 10d action~\eqref{2bw} contribute to $\Delta$ and hence to the 4d curvature. This derivation is very well known for Minkowski compactifications where non-perturbative gaugino condensation and uplift terms (anti D-branes) are absent~\cite{Giddings:2001yu}. In this case, the integrated contributions to $\Delta$ vanish exactly. As before, rather than reproducing the derivation from scratch, we take the setup of~\cite{Giddings:2001yu} as our starting point and consider perturbations induced by gaugino condensation and $\overline{D3}$ branes. These will be assumed to be small.\footnote{Whether a perturbative treatment of such corrections is valid or not in concrete string compactifications is what the dS swampland conjecture questions. We do not attempt to solve this problem in this work.} The corrections would be sub-leading for quantities such as the complex structure of the compactification, which are stabilized in the GKP setup. Hence, we begin with 
\be\label{eq:master}
{\cal V}_6 {\cal R}_\eta= -2\int d^6y\,\sqrt{g} \,\Omega^4 \left(\Delta^{\langle \lambda\lambda \rangle} +\Delta^{\rm other}\right)\,,
\ee
where $\Delta^{\langle \lambda\lambda \rangle}$ correspond to the terms including $\langle \lambda\lambda\rangle$, and $\Delta^{\rm other}$ represents other sources such as $\overline{D3}$-branes.\footnote{
As 
mentioned already in Sect.~\ref{trrg}, different versions of this equation can be obtained by integrating with different powers of the warp factor. The difference becomes relevant only when contributions localized in strongly warped regions (namely $\overline{D3}$-branes) exist and will be important in Sect.~\ref{Sec:uplift_trace-reversed}.}

Notice that we are only interested in the integrated effect of $\Delta$. It contains two terms: The first comes from the external components of the energy-momentum tensor and reads
\be\label{eq:exttensor}
\int d^{10}x\,\sqrt{-G} \,T^\mu_\mu = 4\int d^{10}x\,\sqrt{-G}  \,{\cal L}\,.
\ee
The second involves the trace over the internal components, $T^m_m$, and measures the dependence of the action on the overall scaling of the inner metric:
\be\label{eq:inttensor}
\int d^{10}x\,\sqrt{-G}\, T^m_m = 2\int d^{10}x\,G_{mn}\frac{\delta S}{\delta G_{mn}} = R\frac{\partial S}{\partial R}\,.
\ee
Here $S$ is the 10d action without the Einstein-Hilbert term and, as in Sect.~\ref{trrg}, $R$ denotes the overall scale of $G_{mn}$. Putting these together we obtain
\be\label{eq:deltaformula}
\int d^{10}x\sqrt{-G} \,\Delta= \left({1\over4}R\frac{\partial}{\partial R}-1\right)S \,,\qquad\Longrightarrow\qquad {\cal V}_4{\cal V}_6{\cal R}_\eta=-2\left({1\over4}R\frac{\partial}{\partial R}-1\right)S\,.
\ee
In the following subsections, we compute the 4d curvature taking into account gaugino condensates and $\overline{D3}$ branes.

Before performing the actual computation, our approach to deriving $\Delta^{\langle \lambda\lambda \rangle}$ deserves some discussion. The starting point is the 10d action~\eqref{gte}. The 10d path integral is divided into the closed string sector (which includes gravity and 3-form flux) and the open string one (in particular gauge and gaugino fields). The latter is localized on the D7 brane. We perform the path integral over the open string modes to obtain an effective 10d action for the closed modes. Zero modes of the open strings, which are constant along the brane, are lighter than the KK scale and naturally induce non-local terms in the effective action. The main contribution comes from the non-perturbative dynamics of such zero modes, represented by the exponential dependence of the gaugino bilinear on the K\"ahler modulus $T$ (which is implied by 4d considerations). Schematically,
\begin{eqnarray}\label{eq}
{\cal L}[g_{MN},G_3;\lambda\lambda,F]&\sim& {\cal L}_{\text{closed}}+|G_3 - \lambda\lambda|^2 + {\cal L}_{\text{open}}\nonumber\\
\longrightarrow \qquad {\cal L}_{eff}[g_{MN},G_3] &\sim& {\cal L}_{\text{closed}}\,+\, |G_3 - e^{-T}|^2+\ldots
\end{eqnarray}
The resulting effective action is clearly non-local from the 10d point of view. In particular, it involves a function of $T$, which represents the inner volume. 

While non-local in brane-parallel directions, these effects nevertheless remain localized to the D7 worldvolume. Indeed, microscopically the gauge and gaugino zero modes are short open-string states. The scale relevant for their localization to the brane is the string scale. Moreover, the brane itself has a high tension at weak coupling and hence does not fluctuate transversely.

The leading order terms taken from~\eqref{eq} break the no-scale structure at the order $V_{eff}\sim W_0^2/{\cal V}_6^2$. As usual, we assume that $W_0\sim e^{-T}\ll 1$ can be tuned sufficiently small to maintain perturbative control. One may wonder whether, upon integration over open string modes, other non-local effects competitive with the above can arise. A natural candidate are loop effects, which can however be independently argued to be suppressed (corrections are in general expected to enter at order~$\sim W_0^2/{\cal V}^{10/3}_6$, see refs.~\cite{vonGersdorff:2005bf,Berg:2005ja}). A subtle question is the interplay between loop and non-perturbative effects, i.e. fluctuations of light modes around classical instanton backgrounds. We expect such corrections to be subleading, but their estimation is beyond the scope of our work. At this stage we simply assume that they are negligible. Finally, a subtlety that we have not addressed is the standard assumption that $\langle (\lambda\lambda)^2\rangle$ and $\langle \lambda\lambda\rangle^2$ are of the same order~\cite{Dine:1985rz}.

\subsection{Curvature before the uplifting}
As a first step, let us study an AdS compactification with no $\overline{D3}$ branes, where the 4d curvature arises solely from gaugino condensates. An important point to notice is that, since gaugino contributions are dominated by the unwarped region, factors of $\Omega$ are irrelevant and can be simply set to one. In fact, the parameter $R$ and the standard modulus $T$ are interchangeable in this situation.

The integrated contribution to $\Delta^{\langle \lambda\lambda \rangle}$ can be obtained from the 10d action~\eqref{gte}:\footnote{In what follows, we set the order ${\cal O}$(1) parameter $c=1$ and neglect overall factors.}
\be\label{eq:finiteaction}
S_{\langle \lambda\lambda \rangle}
= {1\over4\kappa_{10}^2}\, \int d^4x \,\sqrt{-\eta} \,\left[
- |\lambda\lambda|^2 T^{1/2}
+\sqrt{g_s}
\left(
T^{1/4}\bar\lambda\bar\lambda\,W_0+\text{c.c.}
\right)\right] \,.
\ee
$W_0$ was defined below \eqref{eq:previous_result}. We can now easily compute the (integrated) contributions to $\Delta^{\langle \lambda\lambda \rangle}$ from~\eqref{eq:deltaformula} (with $R^4\to T$) and read off their effect on the 4d curvature:
\begin{eqnarray}
{\cal V}_6 {\cal R}_\eta&\sim& T\frac{\partial (T^{1/2}|\langle \lambda\lambda \rangle|^2)}{\partial T}
-
\left[\sqrt{g_s} \,W_0\, T\,\frac{\partial(T^{1/4} \langle \bar\lambda\bar\lambda \rangle)}{\partial T}+{\text{c.c.}}
\right]\nonumber\\
&&-T^{1/2}|\langle \lambda\lambda \rangle|^2+\left(\sqrt{g_s}W_0 T^{1/4}\langle \bar\lambda\bar\lambda \rangle+{\text{c.c.}}\right)\,.
\label{eq:AdScurv}
\end{eqnarray}
The first line in this expression contains the contributions from $T^m_m$, as given by~\eqref{eq:inttensor}, while the second comes from $T^\mu_\mu$, eq.~\eqref{eq:exttensor}.

In~\eqref{eq:AdScurv} we have included a dependence of the gaugino bilinear on the Kahler modulus $T$. 
While the introduction of a gaugino condensate inferred from 4d considerations into the 10d action is not a priori justified and may require further scrutiny, our goal is to find whether it leads to consistent results or not. Matching the 4d considerations from previous sections, we hence introduce the gaugino condensate $\langle\lambda\lambda\rangle\simeq g_s^{1/2}  T^{3/4} \,e^{-aT}$,\footnote{The normalization of the gauginos in this section is different from that in Sect.~\ref{Sec:SUSY_AdS}, which was appropriate for the 4d treatment. The relation is $\lambda_{\text{here}}=T^{9/8}\lambda_{\text{Sect.\ref{Sec:SUSY_AdS}}}$.} and $W_0$ is defined below eq.~\eqref{eq:previous_result}. The resulting expression for the curvature is 
\begin{eqnarray}
{\cal V}_6 {\cal R}_\eta&\sim&-\,g_s\,a\,T^2\left(T\,e^{-2aT}-W_0 \,e^{-aT}\right)+\frac{g_s}{2}\,T^2e^{-2aT}\,.
\label{eq:4dgauginocurv}
\end{eqnarray}
Here, as in Sect.~\ref{Sec:SUSY_AdS}, we have taken $W_0$ to be real and positive. This result matches~\eqref{eq:potential_AdS} in a subtle and interesting manner. Notice first that the last term in~\eqref{eq:4dgauginocurv} is subleading in the large volume regime. However, the naively dominant term in brackets vanishes to leading order by the equations of motion~\eqref{eq:TAdS}. In fact, it is very instructive to rewrite~\eqref{eq:4dgauginocurv} as
\begin{eqnarray}
{\cal V}_6 \,{\cal R}_\eta&\sim& \frac{1}{2}\, T^4 \,\frac{\partial V_{\langle \lambda\lambda \rangle}}{\partial T}+T^3 V_{\langle \lambda\lambda \rangle}\nonumber\\
&\stackrel{\text{on-shell}}{\longrightarrow}&T_0^3\, V_{\langle \lambda\lambda \rangle}(T_0)=g_s \,T_0^3\left( {1\over T_0} e^{-2aT_0} - {2\over T_0^2} W_0 \,e^{-aT_0} \right)\,,
\end{eqnarray}
where $V_{\langle \lambda\lambda \rangle}$ is precisely the 4d Einstein frame potential~\eqref{eq:potential_AdS}. Of course, if there are no other contributions to the potential, the first term in the off-shell expression vanishes at the minimum and one recovers the on-shell value of the KKLT curvature, as explicitly written in the second line. Away from this minimum, however, even small shifts of $T_0$  will give a sizeable contribution to the first term coming from the derivatives acting on the exponential factors of $V_{\langle \lambda\lambda \rangle}$. This is a consequence of the steepness of the potential, and is the key ingredient that permits $\overline{D3}$-branes to uplift, as we discuss next.

\subsection{dS uplifting using trace-reversed Einstein equations}\label{Sec:uplift_trace-reversed}
We incorporate next the effect of $\overline{D3}$-branes on warped throats in the previous setup to see how one can in principle obtain dS solutions. The contribution of $\overline{D3}$-branes is, however, much more subtle since they localize in highly warped regions and factors of $\Omega$ become relevant. It is here where version~\eqref{vr2} of the integrated Einstein equations becomes more useful:
\be\label{eq:deltacontributions}
{\cal V}\,{\cal R}_\eta=-2\int d^6y \sqrt{g}\,
\Omega^8\, \Delta^{\langle \lambda\lambda \rangle}-2\int d^6y \sqrt{g}\,
\,\Omega^8 \Delta^{\overline{D3}}\,.
\ee
As we argued, the bulk-dominated pieces in this equation (namely ${\cal V}$ and the gaugino term) are insensitive to which power of the warp factor $\Omega$ is inserted. Hence, they contribute exactly as in~\eqref{eq:4dgauginocurv}. On the other hand, the $\overline{D3}$-brane piece is highly suppressed by a factor of $\Omega^8\ll 1$ and can be simply dropped.\footnote{
Of 
course, one can in principle stick to the integration conventions of~\eqref{vr1}. Although equivalent, this is in practice technically more involved because there arise non-negligible contributions to ${\cal R}_\eta$ from $\overline{D3}$ and the curvature of the warped internal manifold (cf. the discussion at the end of Sect.~\ref{Sec:cancellation}).}

One may naively think that, since the $\overline{D3}$-brane makes no sizeable contribution to the curvature ${\cal R}_{\eta}$, adding it will not uplift the solution to dS. They key point, however, is that the $\overline{D3}$, while not contributing significantly to~\eqref{eq:deltacontributions}, does contribute to the equations of motion. This will induce a small shift in the value of $T_0$ that minimizes the total potential, but as argued before, this has a strong effect in the resulting curvature.

To make this more intuitive, consider a very steep AdS minimum and add an uplifting potential term. One might think that, since the AdS minimum is so steep, the uplift will not shift the value of the field and the contribution to $\Delta$ from the AdS potential will not change. But this is incorrect: On the one hand, the AdS potential becoming steeper does indeed make the field shift smaller. But due to the very same steepness the uplift-effect on the derivative terms involved in $\Delta$ (cf.~e.g.~\eqref{eq:deltaformula}) and hence on the curvature does not disappear. This is how the trace-reversed Einstein equation analysis recovers the simple-mined uplifting logic of the 4d potential approach.

Let us also show this explicitly. As already argued, eq.~\eqref{eq:4dgauginocurv} still follows, even if we now start from~\eqref{eq:deltacontributions}. We can hence still write
\begin{eqnarray}
{\cal V}_6\, {\cal R}_\eta&\sim&-\,g_s\,a\,T^2\left(T\,e^{-2aT}-W_0 \,e^{-aT}\right)+\frac{g_s}{2}\,T^2e^{-2aT}\nonumber\\
&=& \frac{1}{2}\,T^4 \,\frac{\partial V_{\langle \lambda\lambda \rangle}}{\partial T}+T^3 V_{\langle \lambda\lambda \rangle}\,.
\label{eq:upliftcurv}
\end{eqnarray}
As before, we have rewritten the 10d expression (first line) in terms of the gaugino contributions to the 4d Einstein-frame potential~\eqref{eq:potential_AdS}. In contrast to the previous subsection section, however, the first term in the last line no longer vanishes. The on-shell value $T_0$ of the Kahler modulus no longer coincides with the minimum of $V_{\langle \lambda\lambda \rangle}$:
\be
\frac{\partial V_{\text{tot}}}{\partial T}\Big{|}_{T_0}= 0\,\qquad \Longrightarrow\qquad \frac{\partial V_{\langle \lambda\lambda \rangle}}{\partial T}\Big{|}_{T_0}=-\frac{\partial V_{\overline{D3}}}{\partial T}\Big{|}_{T_0}= 4\frac{\mu_3}{T_0^3}\,,
\ee
where $V_{\overline{D3}}(T)$ is the $\overline{D3}$-induced Einstein-frame potential given in~\eqref{eq:potential_dS}. Hence, evaluating~\eqref{eq:upliftcurv} on-shell, we obtain
\begin{eqnarray}
{\cal V}_6 \,{\cal R}_\eta \stackrel{\text{on-shell}}{\longrightarrow} g_s\,T_0^3\left( {1\over T_0} e^{-2aT_0} - {2\over T_0^2} W_0 e^{-aT_0} \right)+2\,\mu_3\,T_0\,.
\end{eqnarray}
This is precisely the 4d KKLT value of the on-shell potential, including the uplift term (in Brans-Dicke frame). 

The equivalence of the 4d and 10d approaches implies that the treatment of the 4d effective potential of KKLT can be reproduced by a 10d analysis. If the appropriate regime of parameters can be obtained in controlled string compactifications, this can lead to dS vacua.
Whether such controlled constructions exist is what the dS swampland conjecture questions, but this analysis shows that they cannot be ruled out simply by the integrated 10d Einstein equation alone.

\section*{Note added}
Work closely related to ours has appeared on the arXiv simultaneously~\cite{Carta:2019rhx,Gautason:2019jwq}. Like our paper, these analyses are concerned with the previously claimed inconsistency of KKLT due to the 10d description of gaugino condensation. Ref.~\cite{Carta:2019rhx} appears to share our conclusion that such a failure of KKLT is not to be expected.\footnote{
They doubt, however, that a fully 10d description can be given and prefer a partially 4d perspective. Independently of this, they also raise a new concern related to the parametric control of the required volume.
} 
By contrast, Ref.~\cite{Gautason:2019jwq} finds different results and claims inconsistency with an uplift to de Sitter.

As far as we can see (cf. also version 2 of~\cite{Gautason:2019jwq}), the main difference comes from the treatment of the gaugino-condensate-dependent piece in the stress tensor appearing in the Einstein equations. We allow the derivative with respect to the external metric components $g_{mn}$ which defines the stress tensor to act on the internal-volume-dependence of the condensate $\langle \lambda\lambda \rangle$ (see~\eqref{vr2} or the $T$-derivative in (\ref{eq:AdScurv})). By contrast, in~\cite{Gautason:2019jwq} the value of  $\langle \lambda\lambda \rangle$ is simply inserted in the previously derived formula for the stress tensor.

Given this disagreement, we want to justify our treatment in some detail (see also \cite{Giddings:2005ff}): We start from the following symbolic form of the full 10d partition function:
\be\label{eq:10d_path_integral}
Z_{10d}=\int {\cal D}{g_{MN}} {\cal D}\phi \,e^{iS[g_{MN},\phi]}=\int {\cal D}{g_{MN}} {\cal D}\phi \,e^{i \int  \left( \frac{1}{2}{\cal R} + {\cal L}(g_{MN}, \phi) \right) }\,.
\ee
Here $\phi$ stands for all fields with the exception of the metric, including specifically gauge fields, gauginos and 3-forms. Integrating out these  matter fields, which is justified in particular because of the mass gap in the confining D7-brane gauge theory, one obtains
\be
Z_{10d} = \int {\cal D}{g_{MN}} \,e^{i\int \frac{1}{2}{\cal R}+iS_{eff}[g_{MN}]}\quad \mbox{with}\quad 
iS_{eff}[g_{MN}]\equiv\ln\left( \int {\cal D}\phi \,e^{i \int {\cal L}(g_{MN}, \phi)}\right). \label{zseff}
\ee
The last expression defines the effective matter action in a standard fashion.

We treat gravity classically, disregarding quantum fluctuations of the metric. The classical solution for $g_{MN}$ is, by definition, the geometry for which the first exponent in (\ref{zseff}) is stationary. This clearly leads to Einstein equations 
\be
{\cal R}_{MN} - {1\over2} {\cal R}\, g_{MN} = T_{MN}^{eff}\qquad \mbox{where}\qquad T_{MN}^{eff}=-{2\over\sqrt{-G}}\,{\delta S_{eff}\over \delta G^{MN}}\,.
\label{ees}
\ee
Now recall that ${\cal L}(g_{MN},\phi)$ contains terms involving gaugino bilinears, e.g. the term $\sim G_3\,\lambda\lambda\,\delta_{D7}$ and our proposed addition $\sim(\lambda\lambda)^2\delta_{D7}$. Following standard lore in SUSY gauge theories, we have
\be
{\int {\cal D}\phi\,\,  \lambda\lambda \,\, e^{i\int {\cal L}}\over \int {\cal D}\phi \,\,e^{i\int {\cal L}}}\,\,=\,\, \langle \lambda\lambda \rangle \,\,\sim\,\, e^{-a  T}.
\ee
Here $\lambda\lambda$ refers to the zero-mode of the D7-brane localized gaugino field and its condensate is hence independent of both internal and external $D7$-brane dimensions. Moreover, consistently with the treatment in~\cite{Dine:1985rz}, we use the approximation
$\langle f(\lambda\lambda)\rangle\,\sim\, f(\langle\lambda\lambda\rangle)\,\sim \, f(e^{-2aT})$. This leads to terms $S_{eff}$ that derive from ${\cal L}$ via the replacement $\lambda\lambda\,\to\,e^{-aT}$. It is now clear that, through (\ref{ees}), these terms enter the Einstein equations exactly according to our approach: Their $T$-dependence and hence their $g_{mn}$-dependence must be used already in the calculation of the stress tensor. Thus we believe that our treatment is appropriate.

Of course, this is only an approximation: Integrating out the confining gauge theory will induce many more terms in $S_{eff}$ than just those arising from $\lambda\lambda\,\to\,e^{-aT}$. This is analogous to integrating out matter fields in Casimir energy calculations, where in addition to the Casimir energy, one generates a series of higher-curvature terms. Similarly, by restricting attention to $G_3 \,\lambda\lambda\,\delta_{D7}$ and $(\lambda\lambda)^2\,\delta_{D7}$, we disregard such higher-order effects in the full 10d brane action.

Before closing this note, it is interesting to think what happens if one inserts the expression for $S_{eff}$ from (\ref{zseff}) in the energy momentum tensor of (\ref{ees}) and exchanges the order of $G^{MN}$-differentiation and functional $\phi$-integration. One obtains
\be
T_{MN}^{eff}=\int {\cal D}\phi \,\,T_{MN}\,\,e^{i\int {\cal L}(g_{MN},\phi)}\,,
\ee
where $T_{MN}$ is the classical stress tensor corresponding to ${\cal L}$. Restricting attention to the terms in $T_{MN}$ which involve the gaugino bilinear 
and replacing $\lambda\lambda$ by $e^{-aT}$ in those corresponds to the treatment advertised in~\cite{Gautason:2019jwq}. But, in our opinion, this is insufficient. Indeed, the fact that the UV gauge coupling (and hence the gaugino condensate) depends on the internal metric $g_{mn}$ is encoded in those terms of $T_{MN}$ which arise come from $\int F_{MN}^2$.

To see the origin of the discrepancy between the two approaches in some more detail, consider a toy model with an action $S=S_0+\delta S$, where
\be
S_0\sim \int d^dx\, \left(F_{\mu\nu}F^{\mu\nu}+\bar{\lambda}\slashed{D}\lambda\right)\qquad\mbox{and}\qquad \delta S\sim \epsilon\, {\cal O}\,.
\ee
Here $\delta S$ is a small perturbation, representing e.g.~the quadratic and quartic gaugino terms in the realistic type IIB model. For example, one can take ${\cal O}\sim T \lambda\lambda$.

The quantity of interest is the trace over internal components of the energy tensor:
\be
T^m_m\,\,\sim\,\, G^{mn}\frac{\delta}{\delta G^{mn}} \log Z \,\,\sim\,\, T\frac{\partial}{\partial T}\log Z\,,
\ee
where $Z$ is the partition function corresponding to $S$.
In our approach, we begin by performing the path integral in the gauge sector. We always work at leading order in $\epsilon$:
\be
\log Z = \log \int {\cal D}A \,{\cal D}\lambda \exp \left(iS_0 +i\epsilon{\cal O}\right) \approx \log \left[Z_0(1+i\langle \epsilon {\cal O}\rangle)\right]\approx \log Z_0 +i\epsilon\langle {\cal O}\rangle\,.
\ee
Here $Z_0$ is the partition function corresponding to $S_0$.
We then take the derivative with respect to the inner metric (or equivalently with respect to $T$). The $Z_0$ factor may be dropped\footnote{Note that $\partial_T Z_0\sim \langle F_{\mu\nu} F^{\mu\nu} + \bar{\lambda}\slashed{D}\lambda\rangle_0=0$, where the subscript indicates evaluation in the (SUSY) theory with just kinetic terms described by $S_0$.}, giving
\be\label{ours}
\frac{\partial}{\partial T} \log Z\approx i\epsilon\,\partial_T \langle {\cal O}\rangle\,.
\ee 
Crucially, $\partial_T$ acts both on any explicit $T$-dependence of ${\cal O}$, but also on the implicit $T$-dependence of expectation values $\langle\ldots \rangle$. In particular, since $T\langle \lambda\lambda\rangle$ depends exponentially on the inverse 4d gauge coupling, i.e. on the volume $T$, the dominant effect at large $T$ comes from the action of $\partial_T$ on $\langle \lambda\lambda\rangle$.

On the other hand, in the approach taken by \cite{Gautason:2019jwq}, one takes the derivative before performing the path integral:
\begin{eqnarray}\label{theirs}
\frac{\partial}{\partial T} \log Z &=& \frac{1}{Z} \int {\cal D}A \,{\cal D}\lambda \left( i\partial_T S_0 +i\epsilon\partial_T {\cal O}\right)e^{iS}\nonumber\\
&=& \frac{1}{Z} \int {\cal D}A \,{\cal D}\lambda \left( iF_{\mu\nu} F^{\mu\nu} + i \bar{\lambda}\slashed{D}\lambda +i\epsilon\partial_T {\cal O}\right)e^{i(S_0+\epsilon {\cal O})}\nonumber\\
&\approx& i\epsilon \langle \partial_T {\cal O}\rangle
\,\, +\,\, i\epsilon \langle \left(iF_{\mu\nu} F^{\mu\nu} + i \bar{\lambda}\slashed{D}\lambda\right) {\cal O}\rangle\end{eqnarray}
The result of~\eqref{ours} does not agree with the first term in this formula. 
As we have argued before, the main contribution to~\eqref{ours} comes from $\partial_T$ acting on the exponential $T$-dependence of the gaugino condensate, which is absent in~\eqref{theirs}. This is precisely the main difference between our results and those of~\cite{Gautason:2019jwq}. However,~\eqref{ours} and~\eqref{theirs} should be equivalent, and so this discrepancy must be compensated by the second term in~\eqref{theirs}. It is not unreasonable to think that suitable non-zero expectation values $\langle F^2{\cal O}\rangle$ and $\langle \bar{\lambda}\slashed{D}\lambda{\cal O}\rangle$ will be present. Their computation is complicated and we view our approach as a simpler method to obtain the desired result.

\section{Conclusions}\label{Sec:conclusion}

Many of the proposed constructions of metastable de Sitter vacua, such as the KKLT scenario \cite{Kachru:2003aw}, involve the interplay of gaugino condensation on a D7-brane stack and an uplift by a positive tension object. As these constructions are defined in the 4d effective field theory, one may question whether there are obstructions in realizing them in 10d where the ultraviolet completion resides. 
Indeed, concerns have been raised using arguments that 
rely on the trace-reversed and integrated 10d Einstein equation 
\cite{Moritz:2017xto, Gautason:2018gln}. 
However, the analyses of brane gaugino condensation in \cite{Moritz:2017xto, Gautason:2018gln}  are plagued by UV divergences, which may prevent one from extracting physically meaningful results.
Recently, we proposed a 10d action for the D7-brane gauginos that is free of such UV divergences \cite{Hamada:2018qef}. The perfect-square structure of the gaugino action also reproduces the 4-fermion terms required by supergravity when compactified to four dimensions (see also \cite{Kallosh:2019oxv}).

Armed with our manifestly finite perfect-square brane gaugino action \cite{Hamada:2018qef}, we give a critical assessment of the concerns raised in \cite{Moritz:2017xto, Gautason:2018gln}.
We first relate an integrated 10d Einstein equation to the extremization condition for a 10d-derived 4d effective potential. The latter can be obtained by dimensionally reducing the 10d action including the perfect-square gaugino term. This effective potential is consistent with 4d supergravity and does not present obstacles for an uplifted minimum. Moreover, within standard approximations, we understand the uplift explicitly in one of the popular versions of the integrated 10d equation. Our conclusion is that de Sitter constructions of the KKLT type cannot be dismissed simply based on the integrated 10d equations considered so far.

Clearly, the 10d approach opens in principle the possibility to go beyond the precision of the standard 4d KKLT analysis. For example, corrections  from D7-brane deformations induced by fluxes and $\overline{D3}$-branes could be calculated. At present, we have nothing to add to the standard arguments that these are subleading. But it is certainly a worthy challenge for the future to develop the 10d approach to a level of precision that would allow for calculating them.

\section*{Acknowledgments}
We would like to thank Ben Freivogel, Daniel Junghans and Thomas Van Riet for useful discussions. The work of YH is supported by the Advanced ERC grant SM-grav, No 669288. 
YH thanks the Laboratoire AstroParticule et Cosmologie (APC) for its hospitality. The work of GS is supported in part by the DOE grant DE-SC0017647 and the Kellett Award of the University of Wisconsin.
We would also like to thank the Simons Center for Geometry and Physics, where this work was initiated during the 2018 Simons Summer workshop, for its hospitality.

\appendix

\section{Hodge decomposition theorem}
The real and complex versions of the Hodge decomposition theorem are
(e.g. Theorem 7.7 and 8.8 in \cite{Nakahara:2003nw}, respectively)
\begin{align}
&\Omega^r(M)=d \Omega^{r-1}(M)\oplus d^\dagger \Omega^{r+1}(M) \oplus {\rm Harm}^r (M),
\\
&\Omega^{r,s}(M)=\partial \Omega^{r-1,s}(M)\oplus \partial^\dagger \Omega^{r+1,s}(M) \oplus {\rm Harm}^r (M),
\end{align}
where $\Omega^r(M)$ and ${\rm Harm}^r (M)$ are the set of $r$-form and harmonic $r$-form on a compact orientable complex manifold $M$ of complex dimension $m$.

We are interested in the decomposition of $\overline{\Omega}_3 \,\delta_{D7}$ of a Calabi-Yau three-fold. The previous theorems guarantee the decomposition
\begin{align}
\overline{\Omega}_3 \,\delta_{D7}
&=d\alpha^{(2)}+d^\dagger\beta^{(4)}+ \gamma^{(3)}
\\
&=\partial^\dagger \tilde{\beta}^{(1,3)}+ \tilde{\gamma} \,\overline{\Omega}_3, \label{eq:decomposition2}
\end{align}
where $X^{(r)}$ represents an $r$-form, and we have used the fact that $\overline{\Omega}_3$ is the unique $(0,3)$-form on a Calabi-Yau threefold. 

We can see that $\partial^\dagger \tilde{\beta}^{(1,3)}$ becomes
\begin{align}
\partial^\dagger \tilde{\beta}^{(1,3)}&=
{1\over2}\left(\partial^\dagger+\overline{\partial}^\dagger\right) \tilde{\beta}^{(1,3)}
+{1\over2}\left(\partial^\dagger-\overline{\partial}^\dagger\right) \tilde{\beta}^{(1,3)}
\nonumber\\
&=
{1\over2}d^\dagger\tilde{\beta}^{(1,3)}+{i\over2}d\left(* \tilde{\beta}^{(1,3)}\right).
\end{align}
Here we have used that $d=\partial+\bar{\partial}$, $d^\dagger=\partial^\dagger+\bar{\partial}^\dagger$, $*:\Omega^{r,s}(M)\to\Omega^{m-s,m-r}(M)$, and the fact that  $(1,2)$-forms are IASD and $(0,3)$-forms are ISD.

Now, from the uniqueness of the decomposition, we can see that
\be
d\alpha^{(2)}={i\over2}d\left(* \tilde{\beta}^{(1,3)}\right),
\quad
d^\dagger\beta^{(4)}={1\over2}d^\dagger\tilde{\beta}^{(1,3)},
\quad
\gamma^{(3)}=\tilde{\gamma} \,\overline{\Omega}_3.
\ee
Therefore, the harmonic part of $\overline{\Omega}_3 \,\delta_{D7}$ is proportional to $\overline{\Omega}_3$.

\bibliography{Bibliography}\bibliographystyle{utphys}

\end{document}